# Optical Rankine Vortex

Grover A. Swartzlander, Jr.

*College of Optical Sciences*
*University of Arizona*
*Tucson, AZ 85718*

and

Raul I. Hernandez-Aranda*

*Photonics and Mathematical Optics Group*
*Tecnologico de Monterrey*
*Monterrey, Mexico 64849*

**Abstract**

Rankine vortex charateristics of a partially coherent optical vortex are explored using classical and physical optics. Unlike a perfectly coherent vortex mode, the circulation is not quantized. Excess circulation is predicted owing to the wave nature of the composite vortex fields. Based on these findings we propose a vortex stellar interferometer.

Vortices exist in all realms of physics and thus the vortex state is universal and robust. In practice the rotational motion is often characterized as a Rankine vortex [1] whereby an inner region near the vortex core rotates as a solid body, and an outer region rotates as an ideal fluid. Within a characteristic radius, $R_{cr}$, the tangential velocity tends to increase linearly with radial distance from the vortex core, and outside this radius it decreases inversely. Optical (and other wavefunction) vortices are frequently described as perfectly coherent modes exhibiting the quantized circulation of an ideal fluid. This Letter describes how spatial coherence properties require an optical vortex to exhibit Rankine vortex characteristics. Not only does the circulation of the system violate the quantization rule, it may also exceed that of the composing fields. This excess circulation is attributed to the wave nature of the system. We first give a geometrical optics account of the system to identify its basic features. Next we briefly introduce a vortex density formalism. Finally we give a statistical optics account that includes numerical computations of ensemble averaged quantities.

Systems described by waves may admit vortex modes when the dimension is two or higher. For a coherent wave the phase of a vortex mode includes a phase factor $\exp(im\theta)$ where $\theta$ is the azimuth measured about the center of the vortex in the transverse $x,y$-plane and $m$ is a non-zero integer called the topological charge. Vortex modes appear in optical systems ranging from the highly symmetric [2]) to the random [3, 4]. Although the characteristics of coherent vortex waves are well understood [5], the nature of partially coherent vortices is still emerging [6-16]. Perfect spatial coherence (and incoherence) is not physical, and thus a lack of understanding is a fundamental problem.

A laboratory system that may be used to generate a partially coherent optical vortex is shown in Fig. 1. An extended uniform source radiating spatially incoherent light is shown. Presently we assume the initial coherence area at the source is negligible. Light propagates through a vortex lens having a surface that resembles the single turn of a helicoid [17,18]. From a geometrical optics point of view, the vortex lens deflects a ray passing through a point ($x_2,y_2$) by an amount $\vec{k}_\theta = m\hat{\theta}_2/r_2$, where $\tan\theta_2 = y_2/x_2$, $r_2 = \sqrt{x_2^2 + y_2^2}$, $\hat{\theta}_2 = -\sin\theta_2\hat{x} + \cos\theta_2\hat{y}$, and $\hat{x}$ and $\hat{y}$ are Cartesian unit vectors. In Fig. 1 the rays 1 and 2 are both deflected in the $\hat{y}$ direction, whereas rays 3 and 4 are respectively deflected in the $\hat{y}$ and $-\hat{y}$ directions. The net azimuthal k-vector at the screen may therefore be expected to vary differently for regions inside and outside the radius $r_3 = a'$, where $a' = az/d$ is the projected radius of the source through the center of the vortex lens. Integrating over all the rays passing through the vortex lens, we find that the net azimuthal k-vector at the screen is given by

$$\vec{k}_{\theta,net}(x_3,y_3) = \frac{\zeta}{\pi a^2}\iint \frac{m}{r_2^2}(-y_2\hat{x} + x_2\hat{y})r_1 dr_1 d\theta_1 = m\zeta(z+d)\hat{\theta}_3 \begin{cases} r_3 d/a^2 z^2, & r_3 < a' \\ 1/r_3 d, & r_3 > a' \end{cases} \quad (2)$$

Thus we find that the geometrical optics description predicts an ideal Rankine vortex whenever the light source has an apparent extent, i.e, $a' > 0$. The scaling factor $\zeta$ is determined by a consideration of the circulation integral: $C = r\int_0^{2\pi} k_\theta d\theta = 2\pi M(r_3)$ where $M(r_3)$ is the vortex strength:

$$M(r_3) = \begin{cases} M_0 r_3^2 / a'^2, & r_3 < a' \\ M_0, & r_3 > a' \end{cases} \quad (3)$$

where $M_0 = m\zeta(1+z/d)$. For $r_3 \gg a'$ we expect $M_0 = m$, and thus we set $\zeta = 1/(1+z/d)$. The transverse coherence length at the vortex lens may be expressed $L_{c,mask} = 0.61\lambda d/a$ where $\lambda$ is the wavelength of the quasi-monochromatic beam [19]. At the coherent limit, $L_{c,mask} \to \infty$, we recover the quantized circulation result across the entire beam: $C_{coh} = 2\pi m$. At the incoherent limit, $L_{c,mask} \to 0$, the circulation vanishes.

By equating the distribution of vortex strength in Eq. (3) with the distribution of topological charge, we interpret the area $r_3 < a'$ as a region having uniform vortex charge density, $m/\pi a'^2$. For $r_3 > a'$ the charge density is equal to zero. This distribution of vortex strength is illustrated in Fig. 2 (uniform gray area), along with the system coordinates. From hereon we drop the subscript on $\vec{r}_3$ for notational convenience.

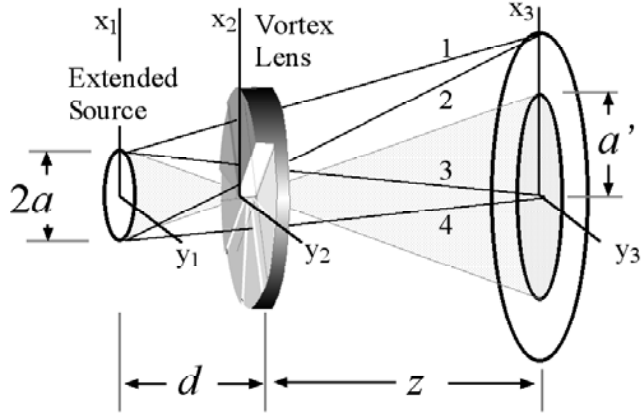

Fig. 1. An extended incoherent light source of radius $a$ illuminates a vortex lens at a distance, $d$. The beam is measured on a screen a distance $z$ behind the lens. Rays 1 and 2 pass through the same region of the lens and are similarly deflected, whereas rays 3 and 4 pass through opposing regions and are counter-deflected. The projection of the light source through the center of the lens demarks a boundary of radius $a'$ between solid body and fluid-like circulation of light.

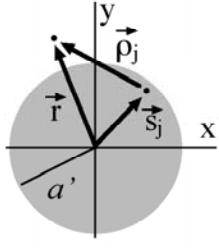

Fig. 2. Gray area of radius $a'$ depicting a uniform region of vortex charge density. The vectors $\vec{r}$ and $\vec{s}_j$, having respective circular coordinates $(r,\theta)$ and $(s_j,\Theta_j)$, indicate locations of a field point and a vortex source point. The difference vector $\vec{\rho}_j$ has circular coordinates $(\rho_j,\psi_j)$.

We now establish a physical optics model of the system that makes use of statistical optics. At any point on the screen the beam is described by an ensemble average of random fields. A member of the ensemble $\{E(r,\theta)\}$ may be constructed from a linear superpositon of vortex fields:

$$E(r,\theta) = \sum_{j=1}^{N} A(\rho_j)\exp[i(m(s_j)\psi_j + \phi_j)] = f(r,\theta)\exp(i\Phi(r,\theta)) \qquad (4)$$

where the points $(r,\theta)$, $(s_j,\Theta_j)$, and $(\rho_j,\psi_j)$, are represented in Fig. 2 by the vectors $\vec{r}$, $\vec{s}_j$, and $\vec{\rho}_j$, respectively. These points are related by the expresssion: $\rho_j\exp(i\psi_j) = r\exp(i\theta) - s_j\exp(i\Theta_j)$. The amplitude $A(\rho_j)$ is an envelope function describing a diffracted vortex mode, $N \gg 1$ is a large integer, and $\phi_j$ is a random variable. A uniform circular distribution of vortices is approximated by setting $m(s_j) = m_0$ within the region $s_j < a'$ and $m(s_j) = 0$ otherwise. Unless stated otherwise, the fundamental case $m_0 = 1$ is assumed. The envelope function is approximated by the expresssion $A(\rho_j) = \tanh(\rho_j/w_v)\exp(-\rho_j^2/w^2)$, where $w_v$ and $w$ respectively characterize the radial size of the vortex core and Gaussian background envelope [20]. Note that when the background is uniform ($w \to \infty$) and $r \gg a', w_v$, the lowest order approximation to Eq. (4) is simply $E(r,\theta) \approx \exp(im_0\theta)\sum_{j=1}^{N}\exp(i\phi_j)$. Far from the optical axis the field in this case may be expected to resemble a vortex having the same charge as the composing fields, $m_0$.

Each member of the ensemble described by Eq. (4) may be assigned real amplitude and phase profiles, $f(r,\theta)$, and $\Phi(r,\theta)$, respectively. Intensity and cross-correlation functions are computed using the expressions:

$$I(r,\theta) = E(r,\theta)E^*(r,\theta) \tag{5a}$$

$$\chi(r,\theta) = \text{Re}\{E(r,\theta)E^*(r,\theta+\pi)\} \tag{5b}$$

Typical examples for one member of the constructed ensemble are depicted in Fig. 3(A-C), showing an off-axis unity charged composite vortex having (A) an amplitude zero, (B) a screw dislocation in the phase, and (C) a ring dislocation in the cross-correlation function. The figures were generated by assigning random values to $\phi_j$, $s_j$, and $\Theta_j$ and coherently summing over $N = 250$ fields. With the aid of a supercomputer the fields were computed on a uniform $1024 \times 1024$ numerical grid with $a'$ spanning 80 grid points. Different members of the ensemble (not shown) were computed using different random seed values.

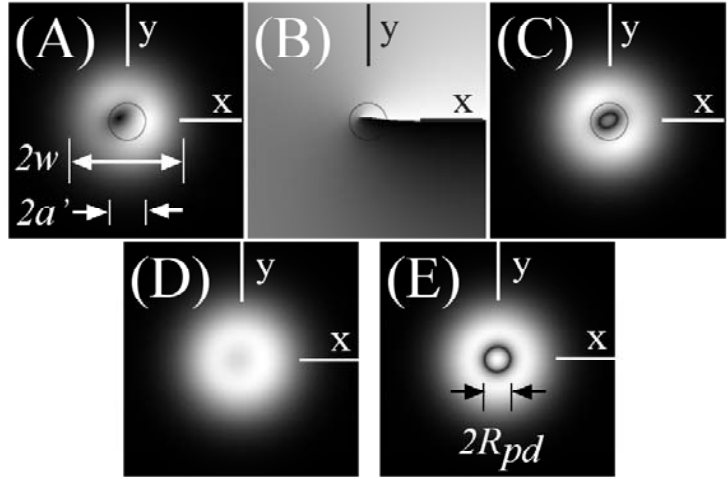

Fig. 3. (A) Amplitude $f(r,\theta)$, (B) phase $\Phi(r,\theta)$, and (C) cross-correlation $\chi(r,\theta)$ profiles for a single member of an ensemble. Ensemble average amplitude (D) and cross-correlation (E) profiles. Each member represents a superposition of 250 vortex fields having a Gaussian envelope of diameter $2w$, fully random phase, and random centroid across a disk of diameter $2a'$. The ensemble is constructed from 300 members.

The centers of the composite vortices (i.e, the points where the intensity vanishes) are found to be randomly distrubuted, as shown in Fig. 4(A). Rather than following a uniform distribution, as assigned for the composing beams (see Fig. 2), we found that the composite vortex locations are Gaussian distributed, as shown in Fig. 4(B). This may be understood as a form of the Law of Large Numbers for composite vortex fields. A least square fit suggests that the standard deviation of the histogram, $\sigma$, may be expressed as $\sigma/w = a'/(a'+w)$. For the example in Fig. 4, a Gaussian envelope having a waist radius of 200 pixels was used for the composing beams. These results did not change significantly when $w_v$ was varied.

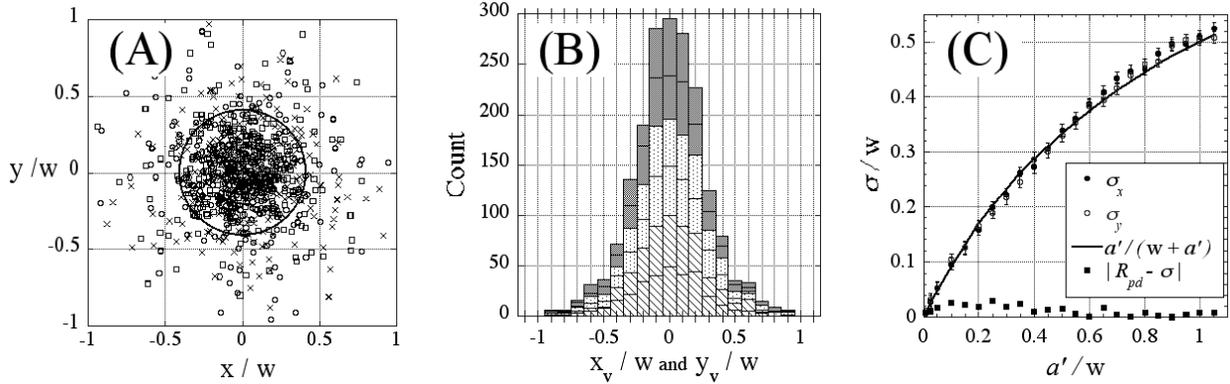

Fig. 4. (A) Scatter plot and (B) histogram of composite vortex positions for 300 members of an ensemble. Both the x and y positions of vortex are represented in (B). The circle in (A) represents the projected source size of radius $a'$. (C) Relative standard deviation of histograms for various values of $a'$. Equivalence of $R_{pd}$ and $\sigma$ is suggested in (C) by the small values of $|R_{pd} - \sigma|$. Least squares fit (line): $R_{pd} \approx \sigma \approx a'/(w+a')$.

Finally we compute ensemble average properties of the beam. The ensemble average intensity and cross-correlation functions are given by $\left\langle \left| E(r,\theta) \right|^2 \right\rangle = \frac{1}{N'} \sum_{l=1}^{N'} I_l(r,\theta)$ and $\left\langle \chi(r,\theta) \right\rangle = \frac{1}{N'} \sum_{l=1}^{N'} \chi_l(r,\theta)$, respectively, where $I_l$ and $\chi_l$ are determined from Eqs. (4) and (5) for the $l^{\text{th}}$ composite field. Examples of these are shown in Fig. 3(D,E) for a 200 pixel Gaussian envelope radius. As expected the low coherence vortex appears as a diffuse gray spot in the center of the beam, rendering the vortex core practically undetectable. In contrast the cross-correlation image contains a distinct high contrast circular phase dislocation of radius $R_{pd}$, as expected [12,13]. As the value of $a'$ increases several things happen: the transverse coherence length $L_c$ decreases (thereby reducing the degree of coherence), the vortex becomes more diffuse, the beam spread widens, and the dislocation radius also tends to increase as $R_{pd}/w = a'/(a'+w)$. This suggests that $R_{pd}$ and $\sigma$ are equivalent. Indeed $|R_{pd} - \sigma| \approx 0$ in Fig. 4(C). Thus, in the opposing limits $a'/w \to 0$ (high spatial coherence) and $a'/w \to \infty$ (low spatial coherence) we find $R_{pd} = \sigma \approx a'$ and $R_{pd} = \sigma \approx w$, respectively.

From an experimental point of view, the relation between $R_{pd}$ and $a'$ allows a determination of the angular extent of the incoherent light source. For a distant nearly unresolvable light source, such as a star, the apparent angular extent of the source, $\Delta\theta = a/d$ may be determined by measuring the diameter of the ring dislocation and the distance from the vortex lens to the measurement screen: $\Delta\theta = z^{-1}/(R_{pd}^{-1} - w^{-1})$. This "vortex stellar interferometer" configuration provides a means of measuring stellar sizes with smaller baselines than the Michelson stellar interferometer if $R_{pd} \ll w$. In the latter case $\Delta\theta \approx R_{pd}/z$.

To explore the relation between the ring phase dislocation in the cross correlation function and Rankine vortex characteristics, we determined the ensemble average topological charge (or vortex strength) enclosed within a circle of radius $r$, $\langle M(r) \rangle$. Values of $\langle M(r) \rangle$ may be computed by first determining the angular spectrum of each member of the ensemble, $f_m(r) = (2\pi)^{-1} \int_0^{2\pi} E(r,\theta) \exp(-im\theta) d\theta$, and averaging the first moments over all members of the ensemble:

$$\langle M(r) \rangle = \left\langle \int_{-\infty}^{\infty} m |f_m(r)|^2 dm \Big/ \int_{-\infty}^{\infty} |f_m(r)|^2 dm \right\rangle \tag{6}$$

Examples for uniform ($w \gg a'$) and Gaussian ($w \approx a'$) background beams are shown in Fig. 5. In the former case $\langle M(r) \rangle$ increases monotonically from zero to the value $m_0 = 1$ (black solid line in Fig. 5). In constrast the Gaussian case exceeds unity before eventually decaying to $m_0 = 1$ (gray solid line in Fig. 5). The enhanced values of $\langle M \rangle > m_0$ are attributed to the high probability of multiple vortices in the composite field in the region beyond the waist ($r > w$), and the subsequent decay is attributed to competition from negatively charged vortices in this region. The spawning of additional vortices is a wave phenomenon attributed to the linear superposition of vortex fields. Corresponding plots (dashed lines in Fig. 5) of the magnitude of the ensemble average azimuthal wave vector, $\langle k_\theta(r)\rangle R_{pd} = \langle M(r)\rangle R_{pd}/r$, exhibit Rankine vortex characteristics and a roll-over point in the vicinity of $r \approx R_{pd}$. The value of the slope of $\langle k_\theta\rangle R_{pd}$ for $r \ll R_{pd}$ in Fig. 5 is approximately 3; this value corresponds to a vortex charge density of roughly unity across an area of $\pi R_{pd}^2$.

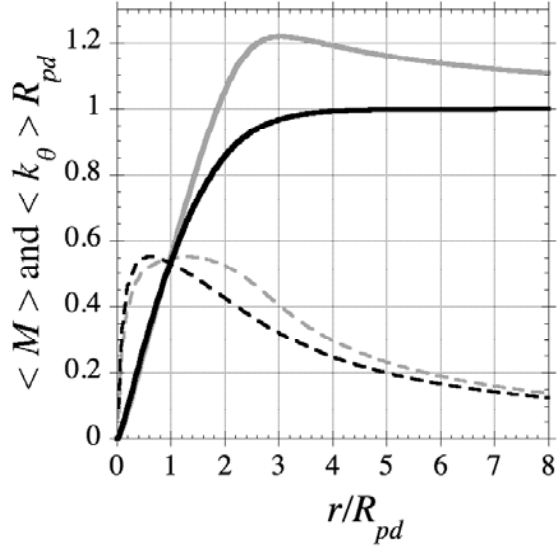

Fig. 5. Solid lines: Ensemble average topological charge, $\langle M(r)\rangle$, as a function of the distance from the optical axis, $r$ (normalized to the phase dislocation radius, $R_{pd}$). Dashed lines: Corresponding normalized ensemble average azimuthal wave vector magnitudes, $\langle k_\theta(r)\rangle R_{pd}$. Black (gray) lines correspond to uniform (Gaussian) background beams.

In conclusion, both geometrical and physical optics models predict Rankine vortex attributes in partially coherent vortex beams. To the extent that perfectly coherent and incoherent light is non-physical, our results suggest that all optical vortex beams must, on average, have a vanishing vortex strength and a vanishing transverse wave vector at the center of the vortex core. In contrast, we note that if $\langle M(r=0)\rangle$ were non-zero, then $\langle k_\theta(r)\rangle$ would non-physically diverge at the origin. The ensemble average vortex charge has been shown to have continuous (non-integer) values, and that these values can exceed the charge of the composing beams. Whereas the ray optics model predicts an abrupt transition from solid-body to fluid-like circulation at the radius of the projected source through the vortex lens, the statistical optics model predicts a smooth transition, which is common in other vortex systems in nature. The radius of this transition is approximately equal to the radius of the phase dislocation of the cross-correlation function. The dislocation radius was found to be less than or equal to the projected source size, depending on the size of the beam envelope. We have suggested using a measurement of the dislocation radius to determine the size of unresolvable an extended object.

This work was supported by the U.S. Army Research Office, the Joint Services Optics Program, Tecnologico de Monterrey (grant CAT007), and CONACYT (grant 42808). The authors acknowledge thoughtful comments by Arvind Marathay (College of Optical Sciences, University of Arizona), Ivan Maleev (KLA Tencor, San Jose, CA) and David Palacios (Jet Propulsion Laboratory, Pasadena, CA).

* Raul Hernandez is a visiting scholar at the College of Optical Sciences, University of Arizona.